\documentclass[aps,prl,twocolumn,showpacs,preprintnumbers,floatfix]{revtex4}
\usepackage{graphicx}

\begin{document}

\title{The resistible effects of Coulomb interaction on nucleus-vapor phase coexistence}

\author{L.~G. Moretto, J.~B. Elliott, L. Phair} 
\affiliation{Nuclear Science Division, Lawrence
        Berkeley National Laboratory, University of
        California, Berkeley, California 94720}

\date{\today}

\begin{abstract}
We explore the effects of Coulomb interaction upon the nuclear liquid vapor phase transition. Because large nuclei ($A>60$) are metastable objects, phases, phase coexistence, and phase transitions cannot be defined with any generality and the analogy to liquid vapor is ill-posed for these heavy systems. However, it is possible to account for the Coulomb interaction in the decay rates and  obtain the coexistence phase diagram for the corresponding uncharged system. 
\end{abstract}

\pacs{21.10.Ft, 
21.65.+f, 
24.10.Pa} 

\preprint{LBNL-53344}
\maketitle

Since the birth of the liquid drop model, which occurred more than 60 years ago, nuclei have been recognized as charged drops of a van der Waals like fluid. Soon after, the concept of cold uncharged, symmetric nuclear matter was introduced. The characterization of its properties, such as its phase diagram and equation of state has been and remains still perhaps the most eminent goal of nuclear physics.

The experimental characterization of cold nuclear matter began by setting the surface, symmetry, and Coulomb terms of the liquid drop expression to zero and retaining just the volume term. This, together with the independent measurement of nuclear radii (already inferable from the surface and Coulomb coefficients), defined the fundamental properties of cold symmetric nuclear matter, namely its binding energy and density at saturation.

Thus, the Coulomb interaction was reasonably eliminated from the picture in order to dispose of what was perceived as a troublesome inessential divergence, while, for better or worse, it remains all pervasive in the experimental realm.

The experimental extension to higher temperatures was hampered both by the lack of a suitable container which became necessary to accommodate the vapor phase (shown to exist by the the soon discovered neutron and proton evaporation) and by the not obviously generalizable finite size effects.

In this paper we consider the problem of the Coulomb interaction and the problem of the container, and we present a simple and natural solution to both. This solution also gives a good indication of how to deal with finite size with maximum generality.

The nuclear liquid vapor phase transition, or more properly, the liquid vapor coexistence is being discussed intensely in literature \cite{Elliott:ISiS, Elliott:EOS2002, Berkenbusch:percolation, Natowitz:limitingT, Gulminelli:critical_behavior, Moretto:CN, Pochodzalla:caloric_curve, Dagostino:neg_C}. Theoretically, a great deal of attention has been given to the effects of the nuclear finite size \cite{Moretto:neg_heat_cap}. Experimentally, discovery and characterization has been claimed from various quarters \cite{Elliott:ISiS, Elliott:EOS2002, Moretto:CN, Berkenbusch:percolation, Natowitz:limitingT, Dagostino:neg_C, Pochodzalla:caloric_curve}, although doubts remain regarding the validity of some approaches and the consistency of the results.

Many of the theoretical approaches have been based upon numerical simulations of finite lattice systems. From these studies negative heat capacities have been claimed to be the signal of the phase transitions in nuclei and other finite systems \cite{Dagostino:neg_C}. From our part we found it more productive to use a simple extension of thermodynamics which incorporates the finiteness of the nuclear system through the surface and other liquid drop terms \cite{Moretto:neg_heat_cap}.  

We showed that a negative heat capacity can be trivially expected in terms of the change of the droplet (nuclear) binding energy with the decreasing size of the evaporating droplet (nucleus). This can be seen most clearly in the case of an isobaric transition:
\begin{equation}
dp = \left. \frac{\partial p}{\partial A}\right| _T dA + \left. \frac{\partial p}{\partial T}\right| _A dT = 0
\end{equation}
or
\begin{equation}
\left. \frac{\partial T}{\partial A}\right| _p = -\frac{\left. \frac{\partial p}{\partial A}\right| _T}{\left. \frac{\partial p}{\partial T}\right| _A}.
\end{equation}
Since from the Clapeyron equation
\begin{equation}
\left. \frac{\partial p}{\partial A}\right| _T\approx -\frac{p}{T}\left. \frac{\partial \Delta H_m}{\partial A}\right|_T
\end{equation}
and 
\begin{equation}
\left. \frac{\partial p}{\partial T}\right| _A \approx p\frac{\Delta H_m}{T^2}
\end{equation}
we can obtain the desired result for the isobaric dependence of the transition temperature on the size of the system
\begin{equation}
\label{eq:partialToverA}
\left. \frac{\partial T}{\partial A}\right|_p = \frac{T}{\Delta H_m}  \left.\frac{\partial \Delta H_m}{\partial A}\right|_T
\end{equation}
or
\begin{equation}
\label{eq:conclusion}
\left. \frac{\partial T}{\partial A}\right|_p = T\left.\frac{\partial\log\Delta H_m}{\partial A}\right|_T .
\end{equation}

In all the above, $A$ is the size of the system, $\Delta H_m$ is the molar enthalpy of vaporization ($\Delta H_m\approx B(A)+T$), and $B(A)$ is the binding energy per particle of a cluster of size $A$.

\begin{figure}
\begin{center}
\includegraphics[width=3.2in]{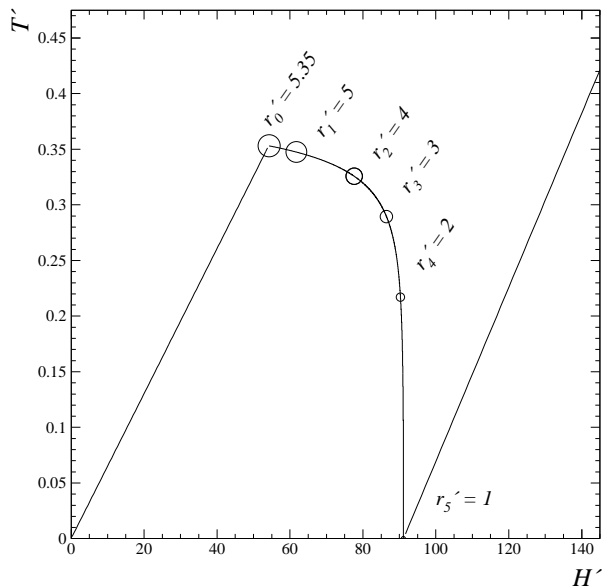}
\caption{The droplet dependent caloric curve at constant pressure. The solid line shows the drop's caloric curve and its dependence on radius. The scaled values of temperature, enthalpy, and radius are $T' = T/\Delta H_{m}^{0}$, $H' = H ( \Delta H^{0}_{m} / 3 c_s
	V_{m}^{l} )^3 / ( 4 \pi / V_{m}^{l} )$ and $r' = r \Delta H_{m}^{0}/(3 c_s V_{m}^{l})$, respectively (see ref.~\cite{Moretto:mesoscopic}). $\Delta H_{m}^{0}$ is the bulk molar enthalpy of vaporization, $c_s$ is the surface energy coefficient, and $V_m^l$ is the molar volume of the liquid.}

\label{fig:neg_Cp}
\end{center}
\end{figure}

In a droplet of a van der Waals liquid, $\Delta H_m$ increases with increasing size $A$ and saturates for the infinite systems. This is of course due to the droplet molar surface energy which decreases and asymptotically vanishes with increasing droplet size. Thus Eq.~(\ref{eq:conclusion}) says that, at fixed pressure $p$ the coexistence temperature decreases as the droplet (nucleus) evaporates. This implies a sloping downwards of the caloric curve in the transition region and a negative heat capacity in the same region. 
An example of the droplet size dependent heat capacity is given in Fig.~\ref{fig:neg_Cp} as the slope in the temperature-enthalpy correlation at constant pressure for different drop sizes. 

However, as we pointed out elsewhere \cite{Moretto:neg_heat_cap}, in nuclei surface effects are not the only ones to be considered. Coulomb effects become progressively more important with increasing $A,Z$ and eventually, at about $A$=60 they reverse the surface trend prevailing at small $A$ values. Above $A\approx$ 60 where the binding is at a maximum the binding and thus $\Delta H_m$ progressively decreases with increasing $A$. The inescapable conclusion is that, within the scope of the Clapeyron equation, no negative heat capacities should be observed for $A>60$, in contradiction with claims to the contrary \cite{Dagostino:neg_C}.

The above result is predicated upon the dominant emission of monomers (neutrons, protons), or, at any rate, very small clusters. Should the nucleus statistically breakup into large fragments, the $A\approx$ 60 upper limit for the observation of negative heat capacities would be correspondingly displaced upwards. However, not only are large fragments rare at low temperature but they never dominate at any temperature (below the critical temperature). In Ising models, finite or not, the average cluster size in the ``gas phase'' hardly rises above unity. The same holds true in experimental data \cite{Elliott:ISiS, Elliott:EOS2002}. It truly seems impossible to escape these thermodynamic conclusions.

However, one may question whether the role of the Coulomb interaction is merely that of decreasing the binding energy. The long range nature of this force may compel us to analyze its role in more detail in first order phase transitions. As will be shown below, the problems are quite serious and threaten our ability to define a true first order phase transition with any generality in the presence of such a force. 

Let us begin with a premise. In typical first order phase ``coexistence'', the two phases (liquid and vapor) do not have to be in actual physical contact. The contact surface is irrelevant as is the short range interaction between the two phases. In other words the equilibrium regime depends exclusively on the properties of each of the two phases, as if the other were not there. Equilibrium exists when the chemical potentials of the two phases are equal, be they in contact or not. 

Let us now introduce the Coulomb interaction in the problem of a drop and its vapor. 

The Coulomb interaction can be split into three parts: 1) the drop self energy; 2) the drop-vapor interaction energy; and 3) the vapor self energy.

The drop self energy, for a finite bound or metastable drop, is easily calculated and does not constitute a problem.

\begin{figure}
\begin{center}
\includegraphics[width=4in, angle=90]{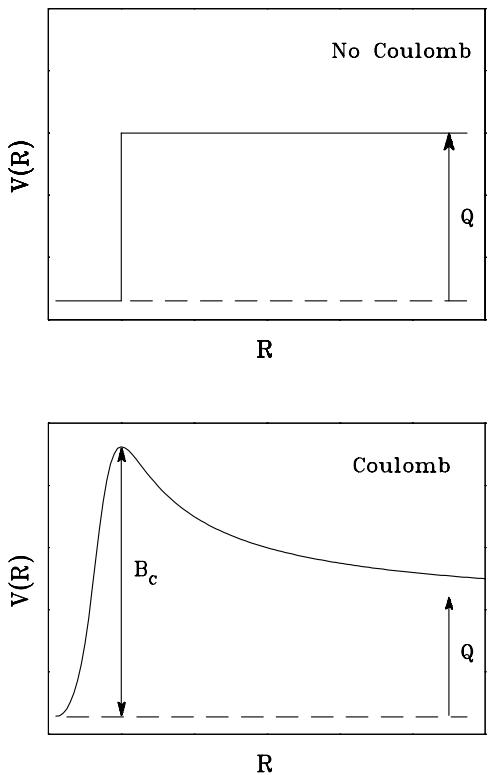}
\caption{Top panel: the schematic potential of an uncharged and bound particle ($Q<0$) leaving a nucleus. Bottom panel: the schematic potential of a bound charged particle. The charged particle must overcome a Coulomb barrier $B_c$ in order to leave the nucleus.}
\label{fig:potential}
\end{center}
\end{figure}

For the drop-vapor interaction, we consider a probe cluster which we can carry from the interior of the drop to infinity.
The potential energy experienced in the process depends upon the particle's charge/mass and is shown schematically in Fig.~\ref{fig:potential}.

If the particle has zero charge (top panel), a step is observed at the droplet radius equal to the particle binding energy. For charges greater than zero, a maximum $B_c$ is observed at the approximate distance of the two droplets in contact. From there the potential decreases according to the Coulomb law and settles down at infinity to a value equal to the binding energy of the particle, $Q$. 

In this case, where we assume that any particle of any size is bound ($Q<0$) and we forget about problem 3), there is no difficulty in defining a gas phase in equilibrium with the droplet \underline {at infinity} constituted by particles of all  sizes whose abundance is controlled by the respective binding energies in the standard way. The intervening Coulomb barrier $B_c$ does not alter the equilibrium, although it may slow its achievement. In this case the vapor is constituted mainly of monomers and the coexistence pressure described by the Clapeyron equation
\begin{equation}
\label{eq:Clapeyron}
\frac{dp}{dT}=\frac{\Delta H_m}{T\Delta V_m}
\end{equation}
with the molar enthalpy $\Delta H_m$ suitably accounting for both surface and Coulomb terms, is completely adequate to describe the liquid to vapor transition and coexistence ($\Delta V_m$ is the molar volume). 

\begin{figure}
\begin{center}
\includegraphics[width=2in, angle=90]{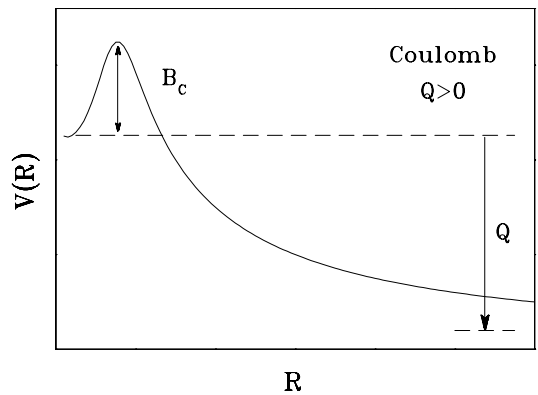}
\caption{The schematic potential of an unbound charged particle.}
\label{fig:unbound}
\end{center}
\end{figure}

Let us now consider the case in which the probe particle becomes unbound to the droplet above some $Z$ value, due to the Coulomb interaction. Now the situation becomes as depicted in Fig.~\ref{fig:unbound}. In this case the droplet is not stable and the ground state of the system may consist of two or more pieces of the original drop at infinity. This is naturally true already at $T=0$. Thus it is not possible to speak properly of this drop in statistical equilibrium with its vapor, since the drop itself is metastable. For a nucleus like gold, the ground state is at least as complicated as three fragments of approximately size 60 nucleons at infinity. This ``true'' ground state is hundreds of MeV below the mass of the gold nucleus. In any statistical calculation, at any reasonable temperature, one can expect a liquid-like phase consisting of a configuration similar to the true ground state in equilibrium with some vapor. A metastable gold-like drop is an immensely improbable configuration because of the great energy chasm mentioned above. The probability of such a configuration can be surmised from the Boltzmann factor $P=\exp(-\Delta E/T)$ where $\Delta E$ is the energy difference between the metastable state and the ground state. Estimating $\Delta E\approx$ 135 MeV and a temperature of 4 MeV we obtain $P\approx e^{-34}$ or approximately $2\times 10^{-15}$.


One might argue that our point is made from energetic rather than free energy considerations and that it may in fact be incorrect. After all, equilibria work both ways, and typically one of the phases is at a lower energy than the other. 

Let us consider, then, the transition from a condensed phase (liquid-like) to a dilute phase (vapor-like). For an infinitesimal isothermal transfer, the variation of the free energy must be zero
\begin{equation}
\Delta F = \Delta E-T\Delta S = 0.
\end{equation}
As we go from liquid to vapor, $\Delta E>0$ for a typical fluid, but this energy increase is compensated by an  equivalent increase in entropy, due just to the increase in molar volume.

However, if $\Delta E$ is negative, due to the Coulomb effect, we need a \underline{decrease} in entropy which is hardly compatible with expansion.

The conclusion is that a statistical equilibrium first order phase coexistence and phase transition is not definable for any droplet that has unbound channels. Of course, the transition of the metastable droplet to its ``true'' complex ground state does not qualify as a statistical phase transition. 

This Coulomb effect seems truly devastating since it does not allow one to define nuclear phase transitions much above $A\approx 60$. 

However there may be a solution to this difficulty. If we consider the emission of particles with a sizable charge, we notice that a large Coulomb barrier $B_c$ is present. For $T<<B_c$ these channels may be considered effectively closed. Consequently the unbound channels may not play a role on a suitably short time scale. Then a phase transition may still be definable in an approximate way. But, of course, we reach again the previous conclusion that for $A>60$ heat capacities must be \underline{positive} and therefore claims of \underline{negative} heat capacities  for large nuclei \cite{Dagostino:neg_C,Chomaz:comment} find here a most serious objection. 


Let us consider now part 3) of the Coulomb energy, namely the vapor self energy. As we said above, it diverges for an infinite amount of vapor. For a dilute vapor, we could consider a small portion such that the intrinsic self energy/nucleon is much less than the temperature $T$.  Alternatively, we could consider a finite box containing a finite system. Unfortunately, at any other distance smaller than infinity the result depends annoyingly on the size (and shape!) of the container and on whether the drop is confined or not in a specified location of the container; a rather inelegant and non-general situation leading to confusing questions about true equilibrium. In any case, it is clear that overall, the Coulomb term makes the definition of phase coexistence and phase transition intractable and ill-posed.

A solution to these difficulties can  be arrived at only by asking a slightly different question: is there a way to obtain experimentally the signal and characterization of the phase diagram (transition) of a nucleus as if the Coulomb interaction were not there? 

As mentioned above, any attempt to define and characterize both phases in the presence of the Coulomb interaction depends (at the very least) on the shape and size of a confining volume applied from without. This seems artificial and lacks a desirable generality. 
But nature actually provides this ``confining volume'' for us. Any particle trying to leave the nucleus is ``boxed in'' by a barrier ($B_s$) which depends on the particle under consideration and on the residual nucleus (or the ``complement''). The top of this potential barrier is close in shape to the potential of two objects, particle and complement, in near contact. The tops of these barriers are actually conditional saddle points \cite{Moretto:famous75}, conditional in the sense that the mass asymmetry is considered frozen. 

According to standard transition state theory all these saddles are in statistical equilibrium with the droplet and the decay rates give direct information on their population which is naturally controlled by a Boltzmann factor $\exp(-B_s/T)$. In particular for large enough $B_s$ the observed experimental abundances are directly related to first chance emission and thus to the transition state rate.

\begin{figure}
\begin{center}
\includegraphics[width=2.4in,angle=90]{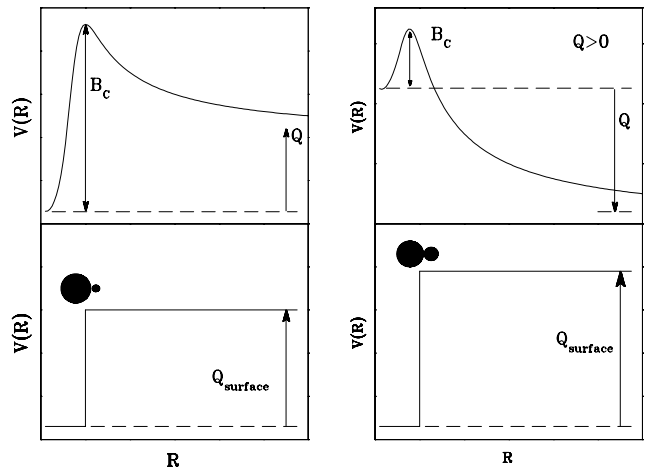}
\caption{A schematic representation of the Coulomb correction when the emitted fragment is bound (left panels) and unbound (right panels). In both cases one can remove the Coulomb energy of the saddle configuration and calculate the $Q$ value using surface energies only (bottom panels). The resulting hypothetical gas will be composed of fragments that are bound to the droplet ($Q_{\rm surface}<0$) for all fragment partitions.}
\label{fig:composite}
\end{center}
\end{figure}

\begin{figure}
\centerline{\includegraphics[width=2.2in,angle=90]{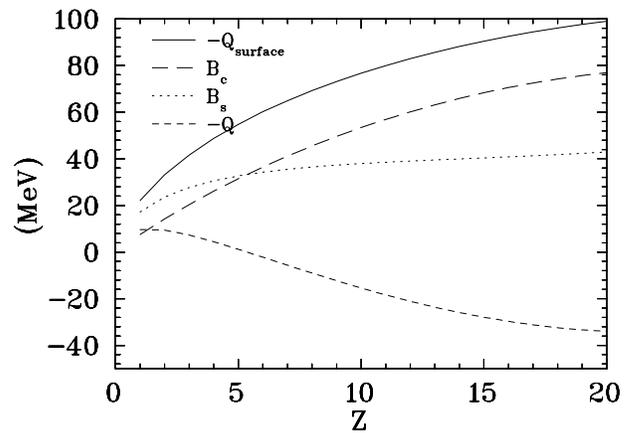}}
\caption{The dependence of the saddle barrier ($B_s$, dotted line) is plotted as a function atomic number for a schematic binding energy which includes only Coulomb and surface terms. Each saddle barrier consist of two parts: the $-Q$ value (short dash) and the Coulomb barrier $B_c$ (long dash). After correction for Coulomb, the relevant Q value calculated using surface terms only binds the fragment to the drop and is shown by the solid line. This is calculated for a $Z$=50, $A$=118 nucleus.}
\label{fig:barriers}
\end{figure}

Now $B_s$ is composed by 
\begin{equation}
B_s=E_{\rm surface}^s-E_{\rm surface}^{gs}+E_{\rm Coul}^s-E_{\rm Coul}^{gs}. 
\end{equation}
where $E_{\rm surface}^s$ and $E_{\rm surface}^{gs}$ are the surface energies of the saddle and ground state respectively, and  $E_{\rm Coul}^s$ and $E_{\rm Coul}^{gs}$ are the Coulomb energies for the same two configurations. Since the Coulomb energies can be easily estimated assuming a two touching spheres configuration for the saddle and one sphere configuration for the droplet (see Fig.~\ref{fig:composite}), we can correct the rates by dividing away the Boltzmann factor containing the Coulomb terms and be left with only the rates/abundances pertaining to the decay of an uncharged drop, for which all channels are bound by the extra surface energy $Q_{\rm surface}=E_{\rm surface}^s-E_{\rm surface}^{gs}$ (see Fig.~\ref{fig:barriers}). 
These rates are now independent of distance and are proportional to the effective partial concentration of the hypothetical gas in equilibrium. We speak of a virtual gas phase because it is not and it need not be present. This picture of a free evaporation of a droplet in vacuum neatly bypasses also the need for a \underline{physical} presence of the vapor. The resulting situation is very much that described by the Fisher droplet model \cite{Fisher:droplet69} for the composition of a saturated vapor in equilibrium with a liquid droplet.
The Fisher droplet model can be directly co-opted to describe the (first chance) fragment abundances of a nuclear physics experiment after correction for Coulomb effects.
From it, it is easy to obtain the coexistence diagram for any nuclear system deprived of the Coulomb interaction \cite{Elliott:ISiS, Elliott:EOS2002}. This is in the same spirit as in nuclear matter calculations in which neutrons and protons are considered as distinct particles, but without any Coulomb interaction.

In summary, we have illustrated the difficulties brought about by the Coulomb interaction in the problem of the nuclear heat capacities. We have shown that for systems with bound decay channels, negative specific heats, if definable at all, are allowed only for $A\lesssim $ 60. For unbound channels, namely nuclei metastable to breakup into two or more large fragments, no phase description or phase coexistence with the normal thermodynamic generality is possible. However, we have shown that decay transition rates can be corrected for the Coulomb interaction in a natural way and the phase diagram for the corresponding chargeless system can be easily obtained.


\bibliography{mybib}

\end{document}